\documentclass[preprint,12pt]{elsarticle}

\usepackage{amssymb}

\journal{Nuclear Physics A  }

\begin{document}

\begin{frontmatter}



\title{Photon and $\eta$ Production in p+Pb and p+C Collisions
  at $\sqrt{s_\mathrm{NN}} = 17.4$~GeV}


\author[addr1]{ M.M.~Aggarwal}
\author[addr2]{Z.~Ahammed}
\author[addr3,deceased]{ A.L.S.~Angelis}
\author[addr4]{ V.~Antonenko}
\author[addr5]{ V.~Arefiev}
\author[addr5]{ V.~Astakhov}
\author[addr5]{ V.~Avdeitchikov}
\author[addr6]{ T.C.~Awes}
\author[addr7]{ P.V.K.S.~Baba}
\author[addr7]{ S.K.~Badyal}
% ...


\address[addr1]{University of Panjab, Chandigarh 160014, India}
\address[addr2]{Variable Energy Cyclotron Centre, Calcutta
   700064, India}
\address[addr3]{University of Geneva, CH-1211 Geneva
   4,Switzerland}
\address[addr4]{RRC ``Kurchatov Institute'',
   RU-123182 Moscow}
\address[addr5]{Joint Institute for Nuclear Research,
   RU-141980 Dubna, Russia}
\address[addr6]{Oak Ridge National
   Laboratory, Oak Ridge, Tennessee 37831-6372, USA}
\address[addr7]{University of Jammu, Jammu 180001, India}
% ...
\address[deceased]{Deceased}

\begin{abstract}
Measurements of direct photon production in p+Pb and p+C collisions at $\sqrt{s_\mathrm{NN}} = 17.4\mathrm{~GeV}$ are presented. Upper limits on the direct photon yield as a function of $p_\mathrm{T}$ are derived and compared to the results for Pb+Pb collisions at $\sqrt{s_\mathrm{NN}} = 17.3$~GeV. The production of the $\eta$ meson, which is an important input to the direct photon signal extraction, has been determined in the $\eta \rightarrow 2\gamma$ channel for p+C collisions at $\sqrt{s_\mathrm{NN}} = 17.4\mathrm{~GeV}$.
\end{abstract}


\end{frontmatter}



The search for a deconfined state of nuclear matter, the so-called Quark-Gluon Plasma (QGP), has been the driving force behind the program of
high-energy heavy-ion research in the last decades.
A variety of results from measurements at the CERN-SPS and RHIC at BNL, and most recently at the LHC at CERN strongly suggest the existence of a deconfined phase.
One of the earliest proposed QGP signatures, the production of thermal direct photons~\cite{Shu78}, has been exceedingly difficult to establish experimentally~\cite{Stankus:2005eq,Reygers:2005sm}.

Since photons do not interact via the strong force, they leave the interaction zone without modification, carrying information about the system at the time of their production. Direct photons, i.e. all photons not originating from decays of long-lived particles, thus allow to probe the entire history of the collision process.
The direct photon yield includes a prompt contribution from initial hard scatterings of the partons, that can be described by perturbative QCD calculations.
Thermal direct photons are those produced in the later equilibrated phase of the heavy-ion collision with a yield that
will be characterized by the temperature of the hot and dense matter created~\cite{Kaj81}.
Measurements of direct photon production in
p+A or p+p collisions are a necessary baseline to determine the prompt photon contribution in A+A collisions. Aside from nuclear effects on parton distribution functions, the prompt direct photon contribution in a nucleus-nucleus collision  is expected to scale with the number of binary nucleon-nucleon collisions, as has been demonstrated in recent measurements by the PHENIX experiment at RHIC~\cite{Adler:2005qk,Adler:2005ig}.
An accurate determination of the prompt direct photon component based on a direct photon measurement in p+p collisions at the same incident energy~\cite{Adler:2005qk}, which is in agreement with pQCD predictions, allowed to extract a possible excess contribution of direct photons in A+A collisions, that may then be attributed to a thermally equilibrated phase.
The spectra of these thermal direct photons can be used to constrain the initial temperature of the medium created in the heavy-ion collision.
Recent measurements of the virtual direct photon spectrum in the electron-positron channel by the PHENIX experiment  have provided first evidence for a direct photon enhancement beyond the expected prompt photon yield in central Au+Au collisions at $\sqrt{s_\mathrm{NN}} = 200$~GeV~\cite{:2008fqa}, which may be explained by a thermal production mechanism. Recent measurements of d+Au collisions at the same center-of-mass energy have been used to improve the understanding of the influence of Cold Nuclear Matter effects on the results from A+A collision~\cite{Sahlmueller:2012ru,Adare:2012vn}.
First results on the direct photon production in heavy-ion collisions at the $\sqrt{s_\mathrm{NN}}=2.76\mathrm{TeV}$ have been reported by the ALICE experiment and are consistent with a thermal contribution to the signal~\cite{Wilde:2012wc}.

The first observation of a significant direct photon signal in heavy-ion collisions was reported by the WA98 experiment~\cite{Aggarwal:2000th,Aggarwal:2000ps} at the SPS.
The direct photon yield was measured in central Pb+Pb collisions at $\sqrt{s_\mathrm{NN}} = 17.3$~GeV. At the time, no direct photon measurement in p+p collisions
at $\sqrt{s} = 17.3$~GeV was available as a reference to determine the expected prompt photon signal.
Instead, direct photon measurements in p+p and p+A collisions at the higher $\sqrt{s_\mathrm{NN}}$ of 19~GeV by other experiments~\cite{prl:mcl83,zpc:bad86,plb:ada95} were used as a reference to determine the prompt photon signal by
applying $x_\mathrm{T}$-scaling ($x_\mathrm{T}=2p_\mathrm{T}/\sqrt{s}$) to account for the difference in the center-of-mass energies~\cite{Aggarwal:2000th,rmp:owe87}. However, these measurements were only available for comparisons above $p_\mathrm{T} \sim 2$ GeV/$c$, and differed from each other by as much as a factor of 2.
Furthermore, pQCD calculations of prompt direct photon production in the SPS energy regime were generally unable to reproduce the measurements without the introduction of intrinsic $k_\mathrm{T}$ contributions~\cite{prc:won98}.  As a result, the prompt photon yield could not be estimated without large systematic uncertainties, and therefore reliable limits on a thermal contribution to the observed direct photon excess in central Pb+Pb collisions could not be inferred without similarly large systematic
uncertainty~\cite{Turbide:2003si,Chat:2009}.
Therefore, the magnitude, and even the existence, of a possible thermal contribution to the direct photon spectrum at SPS energies has remained an open question.

In this paper, results of a direct photon analysis of p+Pb and p+C collisions at $\sqrt{s_\mathrm{NN}} = 17.4$~GeV by the WA98 experiment are presented. Upper limits on the
direct photon yield are obtained and used in a comparison
to the previously published WA98 results on direct photon production in central Pb+Pb collisions at $\sqrt{s_\mathrm{NN}} = 17.3$~GeV.
New measurements of the $\eta$-yield in p+C collisions are presented and combined with the WA98 measurements of the neutral pion yield to extract the $\eta/\pi^{0}$-ratio,  which is a crucial input to the determination of the direct photon yield.

In 1996 the WA98 experiment took data  with a secondary beam of 160~GeV/$c$ momentum on  Pb and C targets,
corresponding to a nucleon-nucleon center-of-mass energy of $\sqrt{s_{NN}}=17.4\mathrm{~GeV}$.
The primary proton beam of 450~GeV/$c$ from the SPS accelerator impinged upon a beryllium target and a secondary mixed beam of 160~GeV/$c$ momentum, consisting primarily of protons and charged pions, was selected and delivered to the WA98 experiment.
A clean beam trigger was defined as a signal in a plastic scintillator start counter,
located 3.3~m upstream of the target, with no coincident signal in a
veto scintillator counter (which had a 3 mm diameter circular hole and
was located 2.7~m upstream of the target), or in beam halo
scintillator counters (covering the transverse region beyond the veto counter up
to a radius of 25~cm).
Two threshold  \v{C}erenkov counters were used for the identification of the beam particles.  Proton projectiles were selected by requiring signals below the corresponding thresholds for pions and kaons in both
detectors. Only events where both  \v{C}erenkov detectors identified the beam particle as below threshold, and therefore a proton, were used in this analysis. The combination of both detectors yields a pion-rejection efficiency of 99$\%$. The minimum bias trigger was
 determined from the total transverse energy deposited in the range $3.5 \leq \eta \leq 5.5$ as measured by the MIRAC calorimeter~\cite{nim:awe89}.

Two targets were used for the data presented here: A $^{208}$Pb target with a thickness $d_{Pb}$ of 0.436~mm and an areal density $\rho_{Pb}$ of 495~mg/cm$^2$, and a $^{12}$C target with $d_{C}=10.022$~mm and $\rho_{C}=1879$~mg/cm$^2$. A total of $1.2 \cdot 10^6$ ($1.0 \cdot 10^6$) minimum bias events were recorded for p+C (p+Pb) collisions, with a factor of $\sim 30$~$(\sim 8)$ more  minimum bias events sampled by a high energy photon trigger described below. The measured WA98 minimum bias cross section $\sigma_{mb}$ of 193~mb (1422~mb) for p+C (p+Pb) corresponds to 86\% (81\%) of the total inelastic cross section.

\begin{figure}[thb]
 \includegraphics[width=0.9\textwidth]{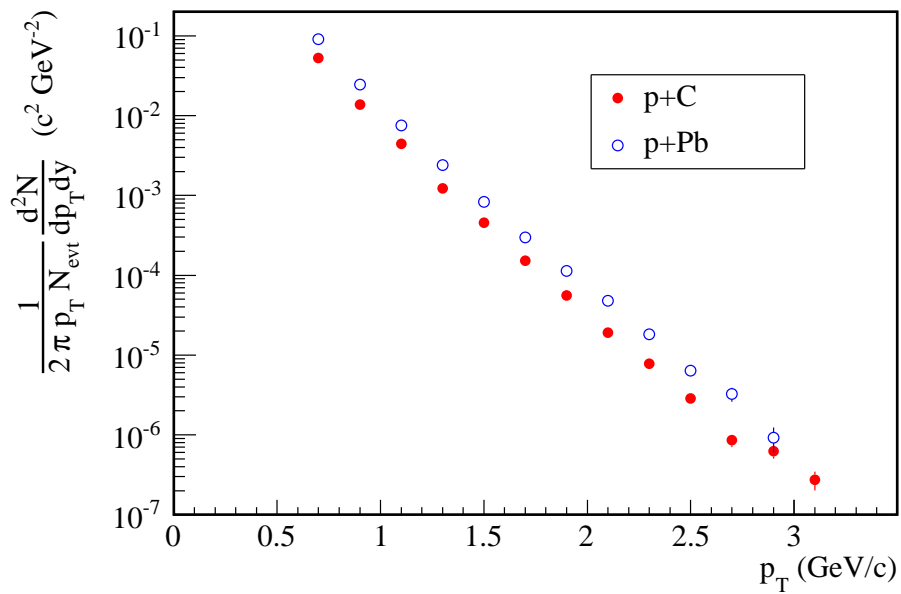}
 \caption{\label{fig:Fig1-pAIncGamma}
 (Color online) Invariant inclusive photon yields from p+Pb and p+C collisions at  $\sqrt{s_\mathrm{NN}} = 17.4$~GeV. The error bars represent the quadratic sum of the statistical and systematic uncertainties.}
\end{figure}

Photons were measured with  the 10,080  module lead-glass electromagnetic  calorimeter (LEDA)  located at a distance of 21.5~m downstream of the target with roughly one quarter of full azimuthal coverage in the pseudo-rapidity range $2.3 \leq \eta \leq 3.0$. A Charged Particle Veto (CPV) detector was placed $1$~m before LEDA to reject charged clusters. To enrich highly energetic photons in the data sample, a high energy photon (HEP) trigger derived from overlapping  $4 \times 4$ groups of adjacent modules (each group offset by 2 towers) in the LEDA was used. The HEP trigger reached its full efficiency for $p_\mathrm{T} \geq 0.8 \mathrm{~GeV}/c$. With the HEP trigger, an additional $1.5 \cdot 10^6$ ($0.5 \cdot 10^6$)  p+C (p+Pb) events were recorded,  equivalent to an additional $3.9 \cdot 10^7$ ($8.2 \cdot 10^6$) sampled minimum bias events.

This analysis follows the procedure described in Ref.~\cite{Aggarwal:2000ps}.
The fully corrected inclusive photon spectra measured with the electromagnetic calorimeter for both the p+Pb and p+C data sets are presented in Fig.~\ref{fig:Fig1-pAIncGamma}. For $p_\mathrm{T} \geq 1.2$~GeV/c the triggered data sample is used.

The spectra have been corrected for the geometrical acceptance and efficiency, analogous to previous measurements by WA98 as described in Refs.~\cite{Aggarwal:2000ps,Aggarwal:2007gw}.
The acceptance correction accounts for the limited spatial coverage of the
detector for single photons. The acceptance is independent of $p_\mathrm{T}$ and amounts to $0.245$ in the rapidity interval $2.0 < y < 3.2$. The efficiency correction has been determined from GEANT simulations in which simulated clusters have been embedded into the raw data to undergo the full reconstruction and analysis procedures that are applied to the measured data. The results have been validated with an independent Monte-Carlo simulation. The efficiency correction accounts for defective detector modules, finite energy resolution effects, and the cuts applied to the data.
The absolute energy scale was fixed to the neutral pion mass as described in Ref.~\cite{Aggarwal:2007gw}.

Off-target events were subtracted on a statistical basis by using the results obtained from data samples taken without a target in place.
The CPV was used to identify photon conversions and determine the fraction of charged particles registered as photons in the LEDA.
As the CPV was not  available during all p+A runs, the fraction of charged particles has been determined on a statistical basis and with a limited $p_\mathrm{T}$ coverage. It has been extended by scaling the correction used in peripheral Pb+Pb collisions, the scaling factors have been validated with Hijing and AMPT simulations.
The correction for the amount of (anti-)neutrons falsely detected in the LEDA as photon hits was determined from simulations based on the results for peripheral Pb+Pb collisions presented in Ref.~\cite{Aggarwal:2000ps}, using scaling factors determined from the same Hijing and AMPT simulations that were used for the CPV information. To account for the steeply falling shape of the spectra, the contents of the $p_\mathrm{T}$ bins are shifted along the y-axis employing a fit with a Hagedorn function $f_\mathrm{Hag} = a \cdot \left[b/(b+p_\mathrm{T})\right]^2$ to represent the correct value at the bin center.

The systematic errors on the inclusive photon measurement are summarized in Table~\ref{tab:WA98IncGammaSysErrors} for two representative $p_\mathrm{T}$ values. The uncertainty of the efficiency correction includes the uncertainties introduced by the cut on the shower shape in the detector, by the finite energy resolution of the detector, which is the dominant uncertainty at the highest $p_\mathrm{T}$, as well as the conversion correction. The $p_\mathrm{T}$ dependent uncertainty for the energy scale has been determined by variation of the scale within its $\pm 1.5 \%$ accuracy, estimated from the neutral pion analysis. The errors bars of Fig.~\ref{fig:Fig1-pAIncGamma} show the quadratically summed statistical and systematic uncertainties

\begin{figure}[thb]
\includegraphics[width=0.9\textwidth]{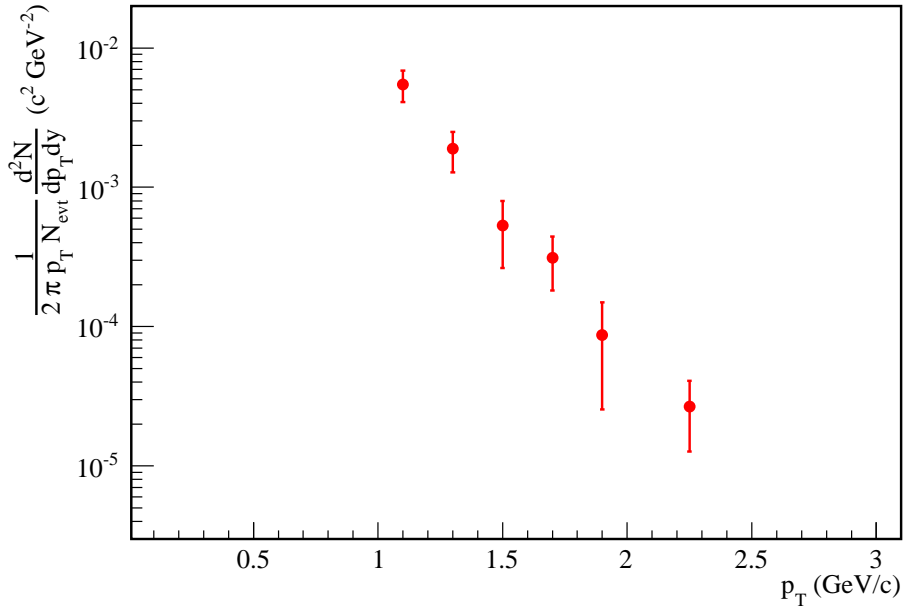}
 \caption{\label{fig:Fig3-pCEtaSpectra}
 (Color online) Invariant $\eta$ yield measured in p+C collisions at  $\sqrt{s_\mathrm{NN}} = 17.4$~GeV in the $\eta \rightarrow 2\gamma$ channel. The error bars represent the quadratic sum of the statistical and systematic uncertainties.}
\end{figure}

The direct photon yield is calculated as a fraction of the inclusive photon yield presented above, following the method described in Ref.~\cite{Aggarwal:2000ps}. The direct photon yield for each $p_\mathrm{T}$ interval is calculated as:
\[
\gamma_\mathrm{direct} = \gamma_\mathrm{inclusive} - \gamma_\mathrm{decay} = \left( 1 - \frac{1}{R_\gamma}\right) \cdot \gamma_\mathrm{inclusive};
\]\[
\qquad R_\gamma = \frac{(\gamma/\pi^0)_\mathrm{meas}}{(\gamma/\pi^0)_\mathrm{decay}}
\]

In the double-ratio $R_\gamma$, systematic uncertainties that are present in the neutral pion and in the inclusive photon measurement partially cancel, thereby reducing the total systematic uncertainty on the measurement. The neutral pion yield in the $\pi^0 \rightarrow 2\gamma$ channel for the two data sets has been presented in Ref.~\cite{Aggarwal:2007gw}. We use a fit to the  $\pi^0$ spectra with a Hagedorn function as reference in the measured $(\gamma/\pi^0)_\mathrm{meas}$ ratio to overcome the different bin widths.

To determine the $(\gamma/\pi^0)_{decay}$ ratio for the decay photons a Monte Carlo simulation was employed to calculate the photonic decays of the relevant mesons -- $\pi^0$, $\eta$, $\eta'$, $\omega$, and $K^0_\mathrm{s}$ -- based on the measured neutral pion spectra in p+Pb (p+C) collisions and by employing $m_\mathrm{T}$-scaling for the heavier mesons. An $\eta/\pi^0$-ratio of $0.55 \pm 0.08$ was assumed~\cite{Aggarwal:2000ps}, consistent with the measured result presented here.

\begin{figure}[thb]
\includegraphics[width=0.9\textwidth]{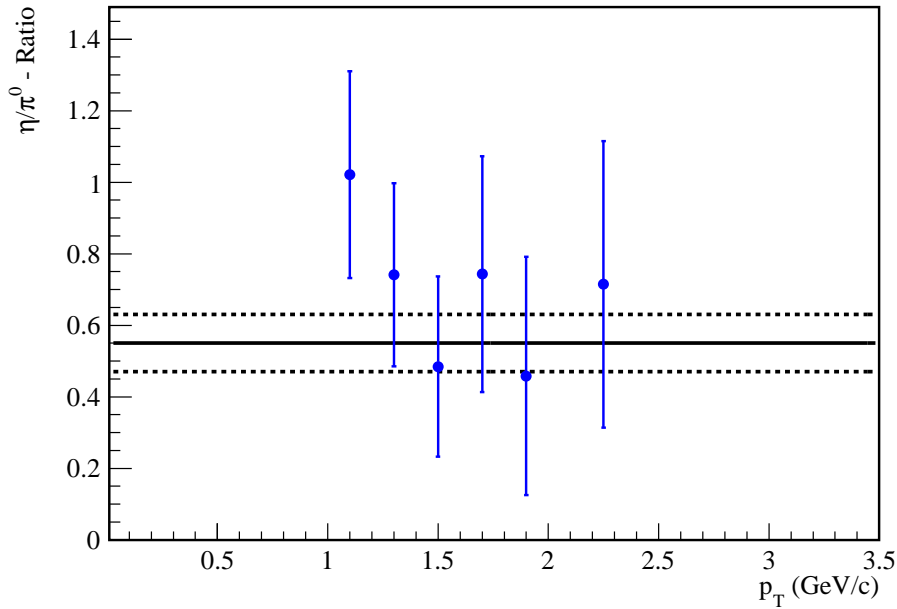}
 \caption{\label{fig:Fig4-eta_zu_pi0}
 (Color online) The ratio of the $\eta$ and $\pi^0$ yield as a function of $p_\mathrm{T}$ in p+C collisions at  $\sqrt{s_\mathrm{NN}} = 17.4$~GeV.
 The $\pi^0$ spectrum is taken from~\cite{Aggarwal:2007gw}. The error bars represent the quadratic sum of the statistical and systematic uncertainties. The ratio assumed in~\cite{Aggarwal:2000ps} with its respective uncertainties is indicated by the horizontal lines. }
\end{figure}

The $\eta$-production in  p+Pb and p+C collisions at $\sqrt{s_\mathrm{NN}} = 17.4$~GeV has been analyzed via
the $\eta \rightarrow 2\gamma$-channel by an invariant mass analysis analogous to that used for the neutral pions~\cite{Aggarwal:2007gw}: the two-photon invariant mass is calculated
for all possible photon pair combinations in an event, and the $\eta$ yield is extracted statistically from the counts in the $\eta$ mass peak.
A requirement that the asymmetry of the photon pairs $\alpha = \frac{E_1 - E_2}{E_1+E_2}$, with $E_{1,2}$ being the energies of photon 1 and 2, should be less than 0.7 was applied to reduce the combinatorial background. In order to remove this background, the invariant mass distributions of uncorrelated photon pairs taken from different events with identical multiplicity were determined. These distributions were scaled to the background level in the real pair distribution for each $p_\mathrm{T}$-interval and subtracted to obtain the $p_\mathrm{T}$-dependent $\eta$-yield.
The  acceptance and efficiency corrections to the $\eta$ yield were determined from the same Monte-Carlo simulations  used for the inclusive photon and neutral pion analyses.

\begin{figure}[thb]
 \includegraphics[width=0.9\textwidth]{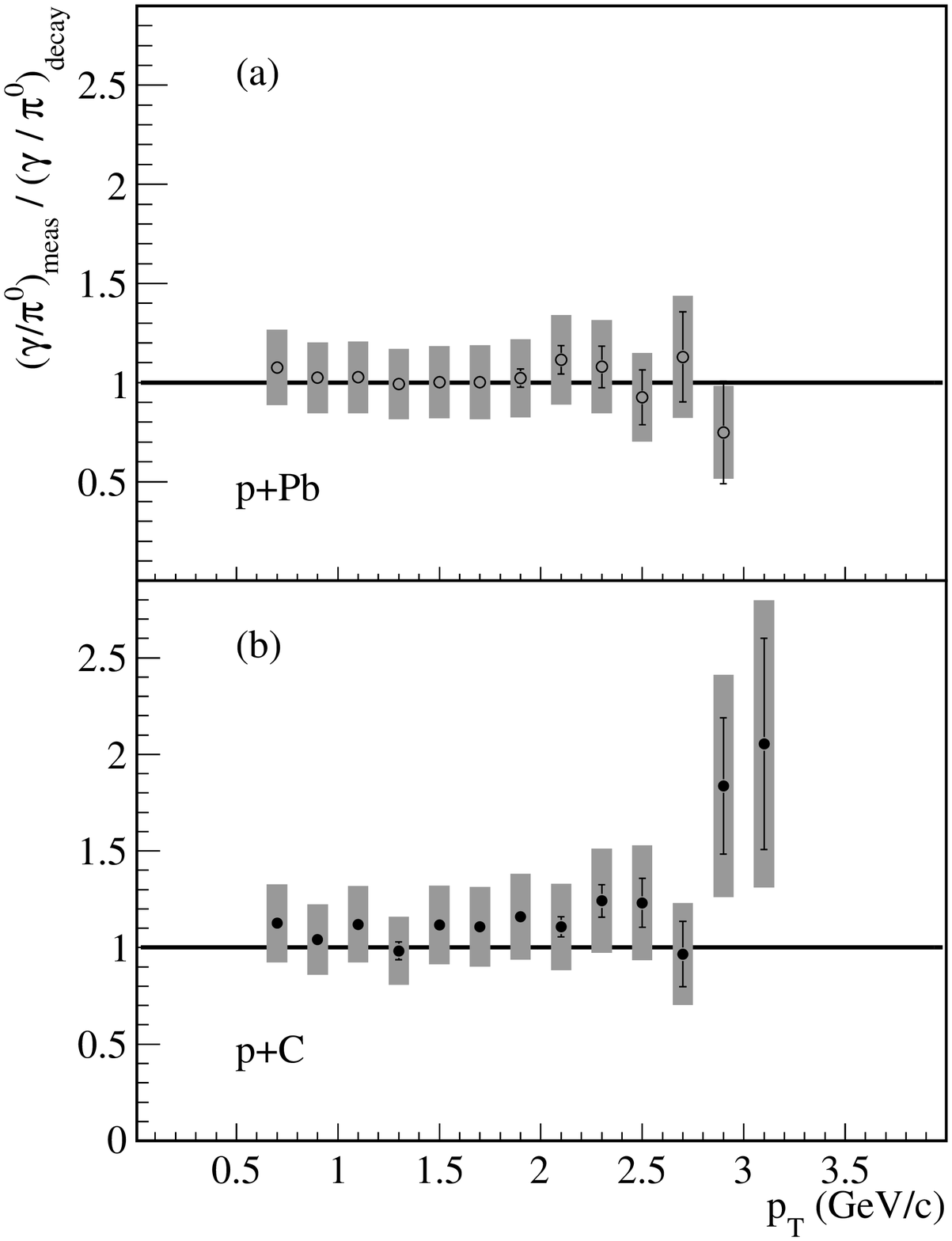}
 \caption{\label{fig:Fig2Ratios}
Double ratio $R_\gamma = (\gamma/\pi^0)_{meas} / (\gamma/\pi^0)_{decay}$  for p+Pb (upper panel) and p+C (lower panel) collisions at  $\sqrt{s_\mathrm{NN}} = 17.4$~GeV. A fit to the neutral pion data from~\cite{Aggarwal:2007gw} has been used for the measured $\pi^0$ result. The error bars show the statistical errors and the shaded areas show the systematic uncertainties of the measurements.}
\end{figure}

A significant $\eta$ result could be obtained only for the p+C data set with the HEP trigger.  The fully corrected $\eta$ spectrum is shown in Fig.~\ref{fig:Fig3-pCEtaSpectra}. The resulting measured $\eta/\pi^0$-ratio for p+C shown in Fig.~\ref{fig:Fig4-eta_zu_pi0} is consistent with the ratio assumed in~\cite{Aggarwal:2000ps}, shown as horizontal lines.

\begin{figure}[thb]
\includegraphics[width=0.9\textwidth]{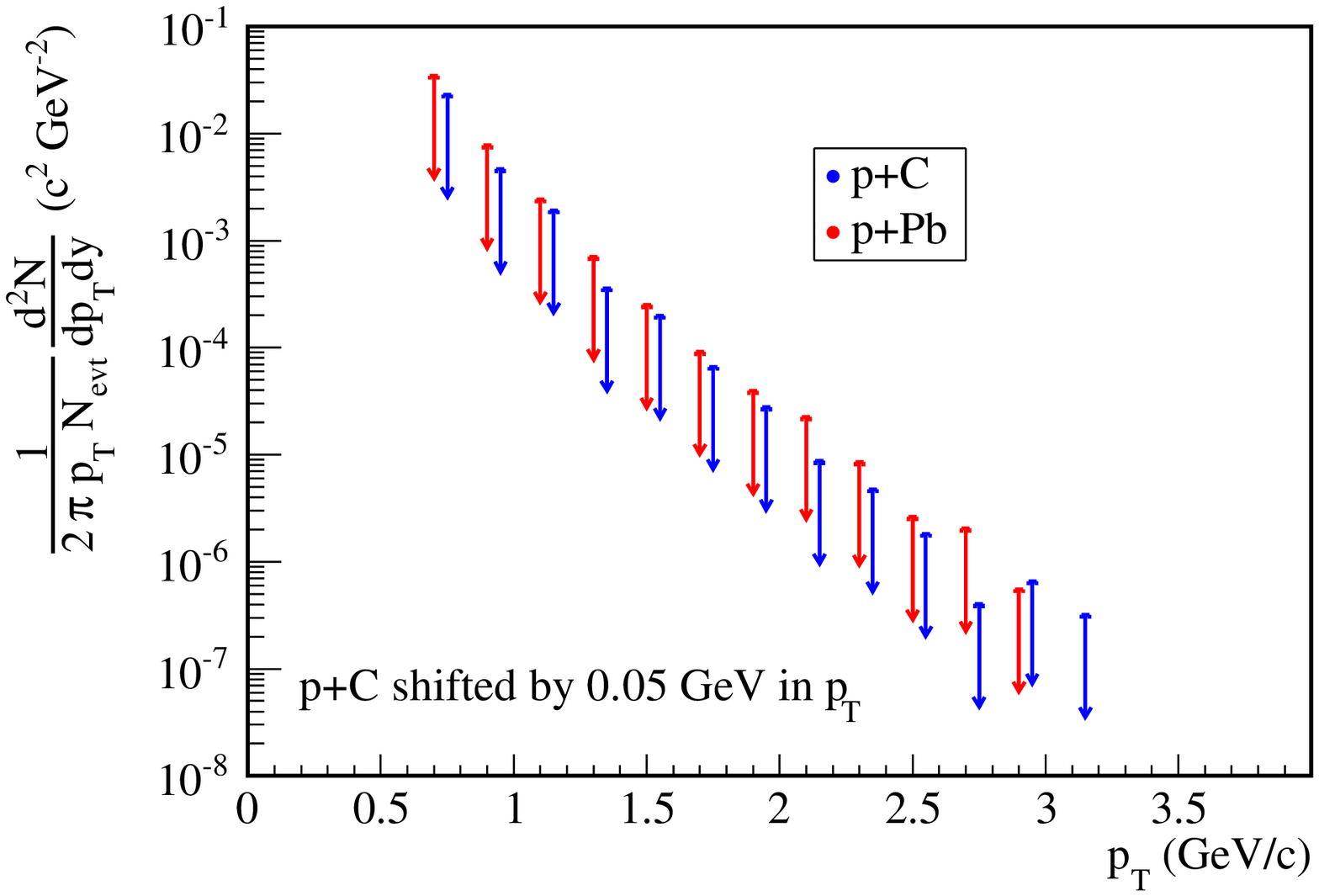}
 \caption{\label{fig:Fig5-DPpASpectra}
 (Color online) Upper limits, quoted at $1.28\sigma$, for the direct photon production in  p+Pb and p+C collisions at  $\sqrt{s_\mathrm{NN}} = 17.4$~GeV. The p+C values have been shifted higher by $0.05$~GeV/c in $p_{T}$ for visibility.}
\end{figure}

The double ratios $R_\gamma$ for p+Pb and p+C are shown in Fig.~\ref{fig:Fig2Ratios}. The uncertainty of the energy scale largely
cancels out in the double ratio.
No significant photon excess is observed in the $p_\mathrm{T}$ range relevant for thermal photon production. However, it is possible to determine upper limits for the production of direct photons up to $p_\mathrm{T}=3.2$~GeV/c, which are shown in Fig.~\ref{fig:Fig5-DPpASpectra}. The upper limits are extracted as $1.28 \sigma$ above the measured result (using $R_{\gamma}= $ max[1,$R_{\gamma}^{meas}$]), as used for the previously published Pb+Pb results~\cite{Aggarwal:2000th,Aggarwal:2000ps,Feldman:1997qc}, and are plotted as vertical arrows with the tail indicating the upper limit.

\begin{table}[ptb]
\begin{center}
\begin{tabular}{|l|c|c|c|}

\hline
      &  $1.25$~$\mathrm{GeV}/c$    & $2.9$~$\mathrm{GeV}/c$  \\
\hline
\hline
Efficiency    & $8.5\%$    &  $21.9\%$   \\
\hline
Background Correction & $2.5\%$    &  $2.5\%$   \\
\hline
Energy Scale &  $13.8\%$  & $28.7\%$ \\
\hline
Geometric Acceptance &  $2.0\%$  & $2.0\%$ \\
\hline
Charged Particles &  $9.2\%$  & $8.0\%$ \\
(Anti-)Neutrons   &  $5.7\%$  & $2.8\%$ \\
\hline
$m_\mathrm{T}$-scaling &  $1\%$   & $1\%$    \\
$\eta/\pi^0$-ratio &   $3\%$  &   $3\%$  \\
\hline
Fit to $\pi^0$-spectrum, p+Pb   &   $2.3\%$    & $7.1\%$ \\
Fit to $\pi^0$-spectrum, p+C   &   $2.4\%$    & $7.0\%$ \\
\hline

\end{tabular}
\end{center}
\caption{\label{tab:WA98IncGammaSysErrors}Systematic uncertainties on the direct photon yield measurement at two representative $p_\mathrm{T}$ values.}
\end{table}

\begin{figure}[thb]
\includegraphics[width=0.9\textwidth]{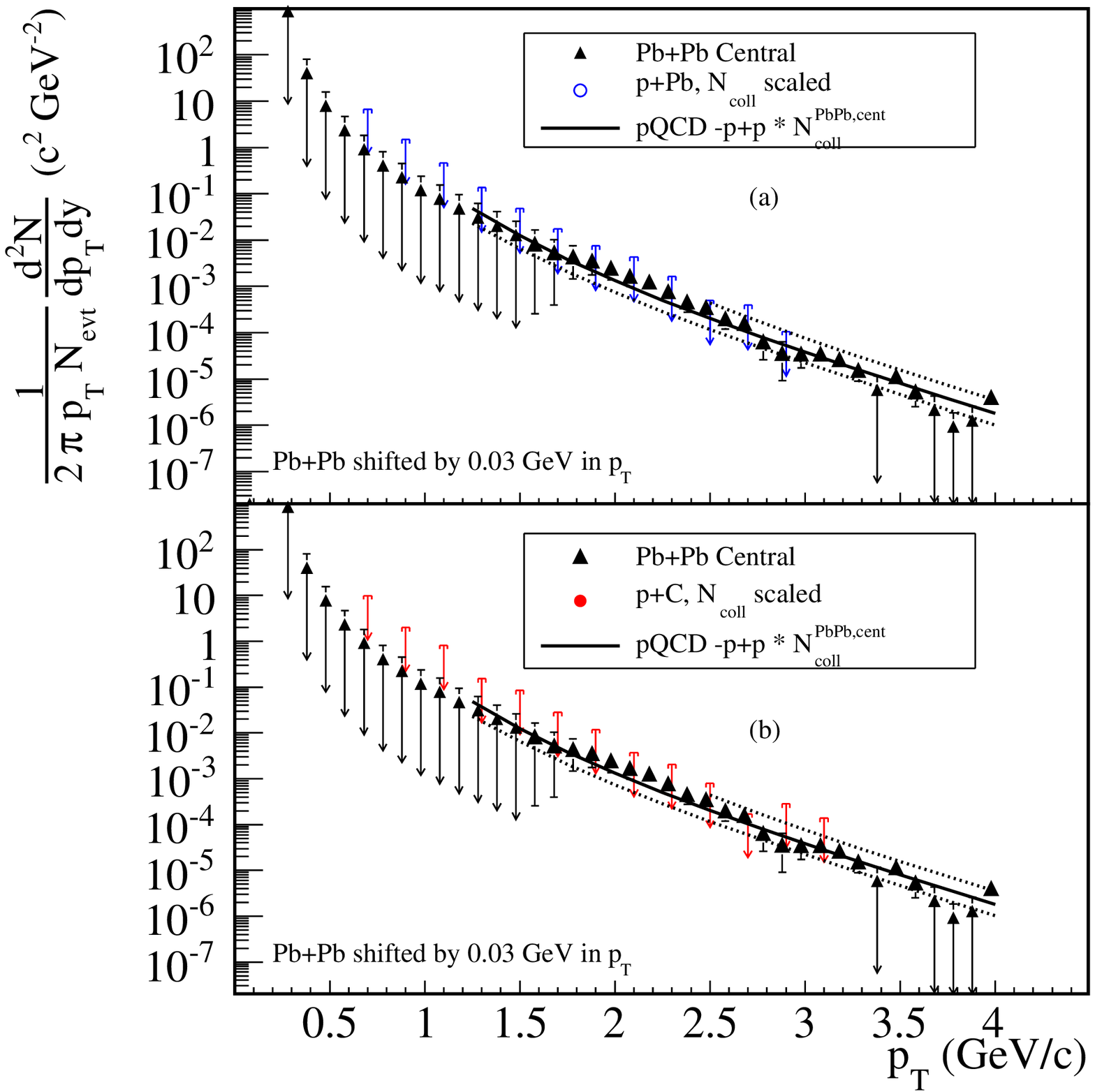}
 \caption{\label{fig:Fig6b_plotPbPbpANLOFrag}
 (Color online) Comparison of the direct photon spectra for central Pb+Pb collision from~\cite{Aggarwal:2000ps} to the p+Pb (upper panel) and p+C results (lower panel) presented in this paper. The two p+A data sets have been scaled by the ratio of the number of binary nucleon-nucleon collisions $N_{coll}^{PbPb}/N_{coll}^{p+A}$. NLO-pQCD calculations for p+p collisions scaled by $N_{coll}^{PbPb}$ for three different scales are shown as lines.}
\end{figure}

These new results are compared to the previously published results for central Pb+Pb collisions~\cite{Aggarwal:2000th,Aggarwal:2000ps} in Fig.~\ref{fig:Fig6b_plotPbPbpANLOFrag} after scaling the upper limits for the p+A datasets by the corresponding ratio of the number of binary nucleon-nucleon collisions, $\left<N_{coll}\right>$, as determined from Glauber calculations~\cite{Miller:2007ri} using a nucleon-nucleon inelastic interaction cross section of $\sigma_{inel}^{NN}=31.8$ mb.
The transverse energy $E_\mathrm{T}$ was modeled by sampling a negative binomial distribution to determine the $E_\mathrm{T}$ contribution of each participating nucleon. To these the same selection on $E_\mathrm{T}$ as used in the event selection by the WA98 trigger was applied, to determine the $\left<N_{coll}\right> $ values. With this procedure the number of binary nucleon-nucleon collisions has been determined as $\left<N_{coll}\right>^{p+C} = 1.7\pm 0.2$ and $\left<N_{coll}\right>^{p+Pb} = 3.8\pm 0.4$~\cite{Aggarwal:2007gw}. For the Pb+Pb data, the events were selected as the 13\% most central fraction of the minimum bias cross section of $\sigma^{Pb+Pb}_{mb} \approx 6300~\mathrm{mb}$, corresponding to $\left<N_{coll}\right>^{Pb+Pb-cent} = 727.8\pm 72.8$.

The direct photon yields observed for central Pb+Pb collisions, scaled by the number of nucleon-nucleon collisions, are consistent with the upper limits from both the p+Pb and p+C results, as shown in Fig.~\ref{fig:Fig6b_plotPbPbpANLOFrag}. Therefore, the new results do not allow to constrain the prompt photon contribution to the Pb+Pb direct photon yield, and hence do not allow to draw definitive conclusions about a possible thermal contribution to the Pb+Pb direct photon yield. A new NLO pQCD calculation is shown in both panels for p+p collisions at $\sqrt{s_\mathrm{NN}}=17.4$~GeV~\cite{deFlorian:2005yj}, scaled by $\left<N_{coll}\right>^{Pb+Pb-cent}$. The calculation employs the GRV photon fragmentation function~\cite{Gluck:1992zx} and is shown for scales $\mu = 0.5p_\mathrm{T}, p_\mathrm{T}, 2p_\mathrm{T}$, which reflect the theoretical uncertainties.  Within these the NLO calculations are able to describe the central Pb+Pb direct photon excess, and hence do not provide evidence for a  thermal direct photon contribution.

In summary, new measurements of $\eta$ production in p+C collisions at $\sqrt{s_\mathrm{NN}} = 17.4$~GeV have been presented. The result is in agreement with  previous  $\eta/\pi^0$-ratio results.  Inclusive photons have been  measured  in p+Pb and p+C collisions at $\sqrt{s_\mathrm{NN}} = 17.4$~GeV, and upper limits on the production of direct photons were obtained. The results cover the $p_\mathrm{T}$ range of $0.7 \mathrm{~GeV}\leq p_\mathrm{T}\leq 3.2\mathrm{~GeV/c}$. Comparison of these upper limits, scaled by $\left<N_{coll}\right>$, to the results obtained by WA98 for central Pb+Pb collisions at $\sqrt{s_\mathrm{NN}} = 17.3$~GeV does not improve the existing experimental constraints on the production of thermal direct photons at SPS energies. Similarly, new NLO pQCD calculations of the prompt photon contribution are consistent, within uncertainties, with the observed central Pb+Pb direct photon excess. Hence, it remains an open question to what extent the direct photon signal observed by WA98 for central Pb+Pb collisions at $\sqrt{s_\mathrm{NN}} = 17.3$~GeV may be attributed to thermal radiation.





\bibliographystyle{model1-num-names}

\end{document}